\documentclass[pre,floatfix,superscriptaddress,twocolumn,showpacs]{revtex4-1}
\usepackage{graphicx}
\usepackage[table]{xcolor}
\usepackage{amsmath,amssymb,amsfonts}
\usepackage{mathtools}
\usepackage{bm}
\usepackage{color}
\usepackage{multirow}
\usepackage{tabularx}
\usepackage{comment}

\begin{document}

\title{Shape matters: Inferring the motility of confluent cells from static images}

\author{Quirine J.~S.~Braat}
\thanks{These authors contributed equally to this work.}
\affiliation{Department of Applied Physics, Eindhoven University of Technology, P.O.~Box 513, 5600 MB Eindhoven, The Netherlands}

\author{Giulia Janzen}
\thanks{These authors contributed equally to this work.}
\affiliation{Department of Applied Physics, Eindhoven University of Technology, P.O.~Box 513, 5600 MB Eindhoven, The Netherlands}
\affiliation{Department of Theoretical Physics, Complutense University of Madrid, 28040 Madrid, Spain}

\author{Bas C.~Jansen} 
\thanks{These authors contributed equally to this work.}
\affiliation{Department of Applied Physics, Eindhoven University of Technology, P.O.~Box 513, 5600 MB Eindhoven, The Netherlands}

\author{Vincent E.~Debets}
\affiliation{Department of Applied Physics, Eindhoven University of Technology, P.O.~Box 513, 5600 MB Eindhoven, The Netherlands}

\author{Simone Ciarella}
\affiliation{Netherlands eScience Center, Amsterdam 1098 XG, The Netherlands}
\affiliation{Laboratoire de Physique de l'Ecole Normale Sup\'erieure, ENS, Universit\'e PSL, CNRS, Sorbonne Universit\'e, Universit\'e de Paris, F-75005 Paris, France}

\author{Liesbeth M.~C.~Janssen}
\email{l.m.c.janssen@tue.nl}

\affiliation{Department of Applied Physics, Eindhoven University of Technology, P.O.~Box 513, 5600 MB Eindhoven, The Netherlands}
\affiliation{Institute for Complex Molecular Systems, Eindhoven University of Technology, P.O.~Box 513, 5600 MB Eindhoven, The Netherlands}

\date\today

\begin{abstract}

Cell motility in dense cell collectives is pivotal in various diseases like cancer metastasis and asthma. A central aspect in these phenomena is the heterogeneity in cell motility, but identifying the motility of individual cells is challenging. Previous work has established the importance of the average cell shape in predicting cell dynamics. Here, we aim to identify the importance of individual cell shape features, rather than collective features, to distinguish between high-motility (active) and low-motility (passive) cells in heterogeneous cell layers. Employing the Cellular Potts Model, we generate simulation snapshots and extract static features as inputs for a simple machine-learning model. Our results show that when the passive cells are non-motile, this machine-learning model can accurately predict whether a cell is passive or active using only single-cell shape features. Furthermore, we explore scenarios where passive cells also exhibit some degree of motility, albeit less than active cells. In such cases, our findings indicate that a neural network trained on shape features can accurately classify cell motility, particularly when the number of active cells is low, and the motility of active cells is significantly higher compared to passive cells. This work offers potential for physics-inspired predictions of single-cell properties with implications for inferring cell dynamics from static histological images.

\end{abstract}

\maketitle

\section{Introduction} \label{sec:intro}

\begin{figure*}
    \centering
    \includegraphics[width=\linewidth]{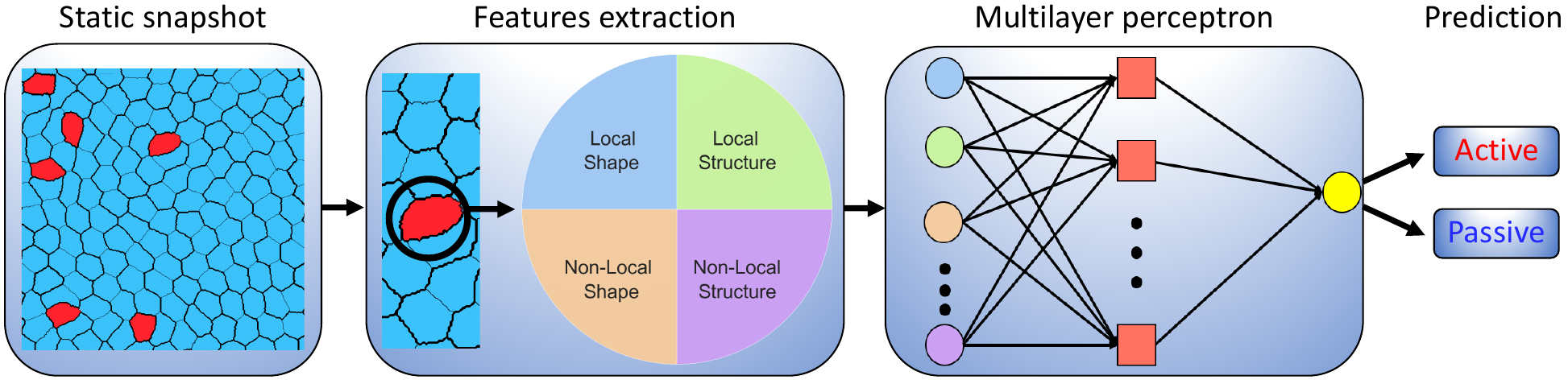}
    \caption{Schematic overview of our machine learning approach for identifying active cells within a mixture of active and passive cells. The Cellular Potts Model generates static cell snapshots, and from each snapshot, a set of shape and structure features is extracted. These features include both local and non-local characteristics. Local features are determined by information on individual cells, while non-local features depend on the cell's neighbours, encompassing neighbour averages, neighbour maximum, and neighbour minimum values. Table \ref{tab:features} shows the complete list of features and the corresponding formulas used to compute them. Local shape features are highlighted in blue, local structure features in green, non-local shape features in orange, and non-local structure features in violet. Following feature extraction, a multilayer perceptron is trained to classify cell types, distinguishing between non-motile (passive) and motile (active) cells solely based on the features extracted from a snapshot. }
    \label{fig:pipeline}
\end{figure*}

Collective cell migration in dense cell layers and tissues is a fundamental process underlying many physiological phenomena including wound healing, embryogenesis, and tissue development, but it also plays a critical role in disease progression such as asthma and cancer~\cite{Stuelten_CellMotilityCancer, Friedl2009}. In general, cell migration is driven by a dynamic interplay between forces, deformations, and environmental cues, making the process complex from both a biological and physical perspective~\cite{Trepat_generalCellMigration, MerinoCasallo_UnravelingCellMigration}. This inherent complexity hampers our ability to reliably predict the migratory capacity of confluent cells and densely packed cellular aggregates, both in healthy and pathological conditions. For prognostic purposes, especially in the context of cancer metastasis \cite{Gottheil_CellUnjamming_Metastasis, Grosser_CellNucleusShape_Fluidity, park2016collective, kas2022cancer}, it would be highly desirable if one could infer information on the expected dynamical behaviour of cellular collectives based solely on \textit{static} information, i.e., from static, microscopic images routinely obtained from histopathology slides. 

Recent breakthroughs have already revealed important morphodynamic links that correlate static, structural features with the collective dynamics of multicellular aggregates. Indeed, pioneering work has established that the average cell shape (as quantified by a dimensionless shape index) in confluent cell layers can serve as a remarkably good proxy for collective cell dynamics, including jamming and unjamming behaviour \cite{Bi_rigidity_nonmotile,Bi_Motility_JammingTransition, Chiang_GlassyInCPM, Czajkowski_HydrodynamicsShapeDrivenTransition, Devanny_Jamming_CPM, Atia2018_geometricconstraint, Park2015, Kim_2020_UnjammingMigration_CancerCell}. Additional static features such as the shape and size of cell nuclei can further refine the predictive power~\cite{Grosser_CellNucleusShape_Fluidity, Gottheil_CellUnjamming_Metastasis, wu2020single}. However, these studies have focused mainly on morphodynamic links for the emergent \textit{collective} cell dynamics. The question to what extent static or structural information can also inform on \textit{single-cell} dynamical properties, such as individual cell motility, has thus far remained largely unexplored. Gaining knowledge about such single-cell properties is particularly important in heterogeneous cell layers, where the presence of more intrinsically motile cells, as in the context of a partial epithelial-to-mesenchymal transition (EMT), is associated with more aggressive cancer progression \cite{Carey_heterogeneous_tumor_exp, Kalluri_basicsofEMT, jolly2019dynamics, Hapach_HeterogeneityMetastasisBreastCancer, Meacham2013_HeterogeneityPlasticity}. 

Here, we seek to derive information about individual cell motility from purely static cell data. In particular, we aim to discriminate between two different cellular phenotypes, high-motility and low-motility cells, based on static images of a minimally heterogeneous in-silico confluent cell layer. The static information that is extracted includes both single-cell geometric shape features and structural properties of the neighbouring cells surrounding a given cell. Our work draws inspiration from Janzen \textit{et al.}~\cite{Janzen_2023_DeathAlive}, who recently investigated the possibility of predicting particle motility in a dense, heterogeneous mixture of spherical active and passive colloidal particles. Briefly, they showed that the shapes of the Voronoi polygons surrounding active particles exhibit distinct characteristics which can serve as sufficient static information to accurately classify different particle motilities. In the present work, we expand upon this approach to study the more challenging, and more biologically realistic, case of a heterogeneous confluent cell layer.

Our confluent cell model is based on the Cellular Potts Model (CPM), a simulation technique that allows for cell-resolved dynamics with controllable single-cell motilities~\cite{Graner_CellSorting, Glazier_Simulation_of_DAdrivenrearrangement, Maree_CPM_book_Hogeweg, Albert_CellShape_MicropatternCPM, VossBohme_2012_CriticalAnalysisCPM, Devanny_Jamming_CPM, Sadhukhan_2021_glassydynamicsCPM_confluenttissue}. To distinguish between high-motility (active) and low-motility (passive) cells, we employ a machine-learning (ML) approach that takes as input instantaneous static information derived from CPM simulation snapshots. Our choice to invoke machine learning stems from the fact that, in recent years, ML has emerged as a powerful tool for identifying structure-dynamics relations in dense disordered passive systems~\cite{PhysRevLett.114.108001, doi:10.1126/science.aai8830, CubukJPC2016, Sussman2018, Boattini2018, Boattini2019, Schoenholz2016, Bapst2020, Paret2020, PhysRevE.101.010602, Boattini2021, Alkemade2022, oyama22, tah22,jung22,ciarella23mlmct,coslovich22,ciarella22tls,Alkemade23,janzen23ageing}, purely active systems~\cite{cichos20,newby18,jeckel19,bo19,munozgil20,tah21,bag21,ruizgarcia22,janzen23ageing}, and active-passive colloidal mixtures~\cite{Janzen_2023_DeathAlive}. Moreover, it has been successfully employed in experimental studies to predict information about the properties of cell collectives \cite{Lyons_CellShape_correlates_metastaticrisk, DOrazio_DecipheringCancerCellBehavior, Kim_2023_PredictionStemCellState_DeepLearning}. 

A schematic overview of our methodology is shown in Fig.~\ref{fig:pipeline}. Briefly, we extract different static features for a given cell from an instantaneous CPM configuration, from which a simple ML algorithm subsequently seeks to classify the cell's motility phenotype. The static input features are subdivided into four categories, namely single-cell (local) shape features, neighbouring-cell (non-local) shape features, local structural features and non-local structural features. The distinction between shape and structural features allows us to identify how much information regarding a cell's intrinsic motility is captured by its shape. To test the validity range of our ML model, we also vary the number of motile cells and their motility strength, thus allowing us to control the cell properties in the heterogeneous confluent layer.

Our analysis reveals that local (single-cell) shape features alone are sufficient to predict whether a cell is highly motile or non-motile. The local shape features work particularly well in the regime where the number of active cells is small and the difference in cell motility between the two cell types is large. In this regime, the cells have a clearly distinct phenotype and local distortions due to a small number of active cells can be more easily detected. These results illustrate that the shape of a single cell contains a significant amount of information about the motility of the individual cell. We also investigate how the ML algorithm performs with different cell parameters and show that the model trained only on local shape features is also successful in generalising to data with a different number of motile cells.

\section{Methods} \label{sec:methods}
\subsection{Simulation model} \label{subsec:CPM}
The ML prediction of a cell's phenotype is derived from  static images produced using the Cellular Potts Model~\cite{Graner_CellSorting, Glazier_Simulation_of_DAdrivenrearrangement}. The CPM is a coarse-grained, lattice-based computational model that simulates cell dynamics via a Monte Carlo algorithm~\cite{Frenkel_ChapterMonteCarlo}. Briefly, cells are represented as pixelated domains on a square lattice, and their dynamics is driven by pixel-copy attempts that minimise the Hamiltonian. We note that recent work has also extended the CPM to disordered lattices~\cite{nemati2024cellularpottsmodeldisordered}. For our study, we utilise the open-source CPM implementation in CompuCell3D~\cite{Swat_CC3D}. 

We simulate a two-dimensional confluent layer composed of both active and passive cells that mimic highly motile and less motile or non-motile cell phenotypes, respectively. The Hamiltonian of our model is defined as follows~\cite{Graner_CellSorting, Glazier_Simulation_of_DAdrivenrearrangement, Guisoni_ModellingActiveCellMovement} 
\begin{equation}
    \begin{split}
    \mathcal{H} & = \mathcal{H}_{adh} +  \mathcal{H}_{area} + \mathcal{H}_{perimeter} + \mathcal{H}_{active} \\
    & = \sum_{\substack{i,j}} J_{\alpha_i, \alpha_j}(1-\delta(\sigma_i,\sigma_j) ) + \sum_\sigma  \lambda_{A} (A_\sigma-A_t)^2 \\
    & \quad\quad + \sum_\sigma  \lambda_{P} (P_\sigma-P_t)^2 - \sum_{\substack{\sigma}} \kappa_\alpha \vec{p}_\sigma \cdot \vec{R}.     
    \end{split}
    \label{eq:general_Hamiltonian}
\end{equation}
The individual pixels are indicated with $i, j$. All cells can be identified with their cell number $\sigma$ and have an associated cell type $\alpha$ (either active or passive). Each of the terms in the Hamiltonian corresponds a different physical aspect of the cells. The first term, $\mathcal{H}_{adh}$, accounts for the change in adhesion energy associated with cell-cell adhesion contacts. The magnitude of the cell-cell adhesion term between the cell type is set by $J_{\alpha_i, \alpha_j}$. The Kronecker delta function ($\delta(\sigma_i, \sigma_j)$) ensures that cells do not experience adhesion interactions with themselves. The second term, $\mathcal{H}_{area}$, penalises large differences between a cell's actual area $A_{\sigma}$ and its preferred area $A_t$ and maintains a cell's size. Similar to the area constraint, an energy penalty term is included for large variations of a cell's perimeter, $\mathcal{H}_{perimeter}$. Contrary to the area constraint, this penalty is only accounted for if the cell's perimeter $P_\sigma$ exceeds the value $P_t$. We include the perimeter constraint to avoid cell shapes with non-physically large perimeters. When a cell's perimeter  is below the threshold $P_t$, no perimeter constraint is applied such that the term does not affect the emerging cell shapes.  Lastly, the motility of the cells is implemented in the term $\mathcal{H}_{active}$. The strength of the motility of the cells is given by $\kappa_\alpha$ and depends on the specific cell type $\alpha$ (either active or passive) for each individual cell. The vector $\vec{p}_{\sigma}$ represents the directional persistence of the cell. Together, these four energy contributions determine the overall dynamics of the cells in the simulations.

Phenotypic heterogeneity is included via the fourth term in the Hamiltonian, i.e.\ the term describing the motility of the cells. In biology, motility is controlled by many intrinsic and external factors~\cite{Trepat_generalCellMigration, SenGupta_PrinciplesCellMigration, MerinoCasallo_UnravelingCellMigration}. In our computational model, we reduce this complexity by using a simple active force implementation that effectively biases the migration direction of the cells. This direction, defined by $\vec{p}_\sigma$, is described by a random orientation that gets updated every Monte Carlo Step (MCS) with a random angular perturbation $\eta \in [-\frac{\pi}{36}, \frac{\pi}{36}]$. The key difference between the active and passive cells is the strength of the active force $\kappa_\alpha$ (which depending on the cell type $\alpha$ is either active (a) or passive (p)). We distinguish between two different scenarios in the simulations, namely 
\begin{enumerate}
    \item the passive cells are non-motile ($\kappa_p$~=~0) and the active cells are motile ($\kappa_a~=~1500$);
    \item both the active and passive cells are motile, but the active cells are more motile ($\kappa_a~>~\kappa_p>0$).
\end{enumerate}
The first situation allows us to investigate how active cells distort the cellular arrangements in a passive cellular environment. The second resembles a more realistic representation of confluent cell layers, as the motility of cells shows heterogeneity even within confluent tissue~\cite{Angelini_GlasslikeDynamicsOfCollectiveCellMigration}. 

\begin{figure}
    \centering
    \includegraphics[width=\linewidth]{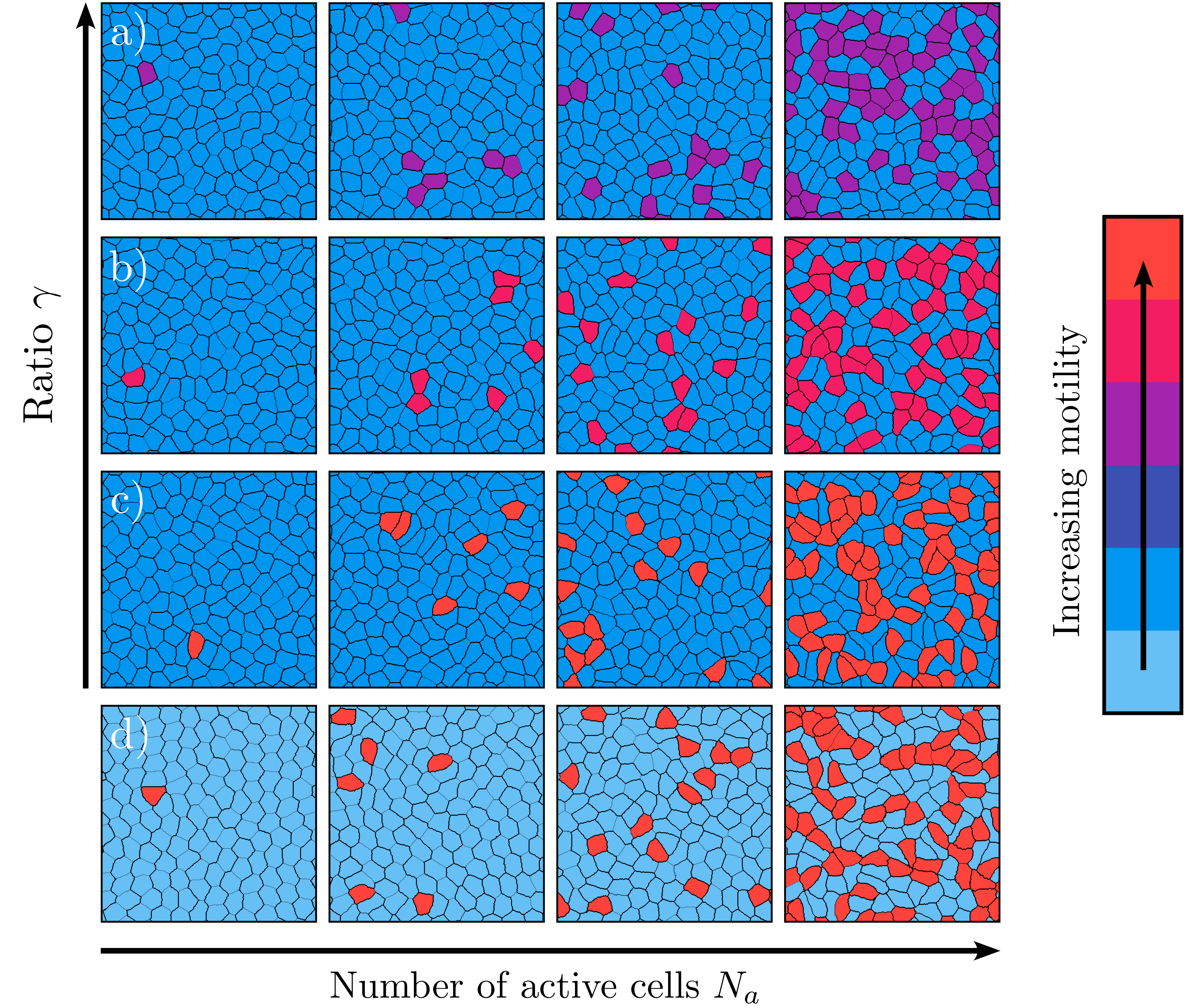} 
    \caption{Overview of the static images extracted from the Cellular Potts simulations. The colours indicate the motility of the different cell types. (a-c) The snapshots in which the number of active cells $N_a$ and the motility of active cells $\kappa_a$ is varied for constant passive cell motility, $\kappa_p~=~150$. The ratio $\gamma$ is defined as $\kappa_p/\kappa_a$, (d) the snapshots in which the number of active cells $N_a$ is varied for constant cell motility, $\kappa_p~=~0$ and $\kappa_a~=~1500$.}
    \label{fig:overview_all}
\end{figure}

For the numerical implementation, we employ a two-dimensional square lattice of $300$~by~$300$~pixels. The simulation contains 144~cells where a number $N_a$ of these cells are randomly chosen to be active, creating a mixture of active and passive cells. We vary the number of active cells between $0$ and $60$. We set the adhesion strength $J_{\alpha_i, \alpha_j}=5.0$ for all cells, and each cell has a target volume $A_t$ of $625$ pixels which is enforced with an energy penalty constraint of $\lambda_A~=~1.0$. To avoid any cell fragmentation, the pixels of an individual cells are forced to remain connected throughout the entire simulation. This can cause artefacts in the cell shapes (long tails are formed). To circumvent this problem, the perimeter constraint (with $\lambda_P~=~1.0$) is applied when the cell perimeter exceeds a value of $P_t~=~150$ pixels.

After equilibration, the static snapshots are stored every $1000$~mcs. This time interval is chosen such that the active cells can move sufficiently between consecutive snapshots.
Snapshots for different parameters are shown in Fig.~\ref{fig:overview_all}. It is challenging to distinguish between the active and passive cells in the static images by visual inspection only. We therefore extract physical features from these snapshots to determine whether a machine-learning algorithm can predict the phenotype of the cell based on a set of simple physical properties. 

\subsection{Classification model} \label{sec:ML}
We approach the task of identifying active cells as a binary classification problem. To accomplish this, we employ a multilayer perceptron~\cite{GARDNER19982627,Multilayer159058}, as implemented in Scikit-learn~\cite{scikit-learn}, which consists of interconnected neurons in multiple layers. The first layer, i.e.\ the input layer, receives the input vector, while the output layer provides output signals or classifications with assigned weights. The hidden layers adjust the weights until the neural network's margin of error is minimised~\cite{haykin2009neural}. In this work, we use a simple neural network with a single hidden layer containing a number of nodes equal to the input features. We update the weights using the ADAM algorithm~\cite{kingma2014adam}.

To evaluate the model's performance, we calculate the accuracy, defined as the number of correct predictions divided by the total predictions. Here, a prediction is a single classification attempt (active or passive) based on the static input features for one given cell from one simulation snapshot (detailed in the section below). When the number of active cells $N_a$ deviates from the number of passive cells, indicating an imbalanced dataset, we address this by randomly selecting a subset of passive cells and excluding them. This method ensures a balanced dataset with the same active and passive cells ($N_a$). We use multiple independent snapshots to obtain a total of $120 \, 000$ cells. Note that the number of snapshots used depends on $N_a$, but the overall number of cells remains fixed. We randomly divide the dataset into training and test sets, allocating $80\%$ of the data to the training set and $20\%$ to the test set. 
We train 20 independent neural networks, and the reported accuracy is the average accuracy obtained from these neural networks. Consequently, while the single-cell features used for training are extracted from multiple snapshots, the trained model can be tested on features extracted from a single snapshot.

While this paper presents results based on the application of a multilayer perceptron, we have confirmed that similar results can be achieved using a more sophisticated ML algorithm, specifically a gradient-boosting model, which is a machine-learning method based on decision trees~\cite{ke2017lightgbm}. Additionally, our results showed that a simplistic model like linear regression~\cite{Bishop2006,mahmoud2019parametric,Tokuda2020} fails to accurately predict whether a cell is active or passive. Consequently, we can conclude that, for this classification problem, a more advanced non-linear model such as a multilayer perceptron is necessary.

\subsection{Input features} \label{sec:input_features}
\renewcommand{\arraystretch}{1.55}
\begin{table*}
    \centering
    \caption{Features employed for the machine learning model with their corresponding formulas. The colour-coded distinctions represent the four feature subsets: local shape features in blue, local structure features in green, non-local shape features in orange, and non-local structure features in violet. The 'boundary pixels' of a cell are defined as pixels with at least one first-order neighbouring pixel belonging to a different cell.}
    \label{tab:features}
    \begin{tabular}{ | m{5cm} | m{7.5cm} | m{1.2cm} | m{1.4cm} | m{1.4cm} |}
    \hline
    \multirow{4}{*}{Parameter} & \multirow{4}{*}{Formula} & \multicolumn{3}{|c|}{Neighbouring Cells}\\
    \cline{3%
    -%
    5}%
    &&\multirow{3}{*}{\rotatebox[origin=c]{90}{Avg.}} & \multirow{3}{*}{\rotatebox[origin=c]{90}{Max.}} & \multirow{3}{*}{\rotatebox[origin=c]{90}{Min.}}\\
    &&&&\\
    &&&&\\
    \hline
    \hline
    Volume (\#pixels) & V \cellcolor{cyan!20}& $\langle V \rangle$ \cellcolor{orange!10}& $V_{\textrm{max}}$ \cellcolor{orange!10}&\cellcolor{orange!10} $V_{\textrm{min}}$\\ \hline
    Surface (\#boundary pixels) & A \cellcolor{cyan!20}&\cellcolor{orange!10} $\langle A \rangle$ \cellcolor{orange!10}&\cellcolor{orange!10} $A_{\textrm{max}}$ & $A_{\textrm{min}}$ \cellcolor{orange!10}\\ \hline
    Surface volume ratio & $\gamma=A/V$ \cellcolor{cyan!20}& $\langle \gamma \rangle$ \cellcolor{orange!10}& $\gamma_{\textrm{max}}$ \cellcolor{orange!10}&\cellcolor{orange!10} $\gamma_{\textrm{min}}$\\ \hline
    Number of neighbours & N \cellcolor{cyan!20}& \cellcolor{gray!25}& \cellcolor{gray!25}& \cellcolor{gray!25}\\ \hline
    $1^{st}$ Moment of mass & $\overline{|r|}=\frac{1}{V}\sum_i^V |\vec{r_i}|$ \cellcolor{cyan!20}& $\langle\overline{|r|}\rangle$ \cellcolor{orange!10}& $\overline{|r|}_{\textrm{max}}$ \cellcolor{orange!10}&\cellcolor{orange!10} $\overline{|r|}_{\textrm{min}}$\\ \hline
    $2^{nd}$ Moment of mass & $\overline{|r|^2}=\frac{1}{V}\sum_i^V |\vec{r_i}|^2$ \cellcolor{cyan!20}&\cellcolor{orange!10} $\langle\overline{|r|^2}\rangle$ & $\overline{|r|^2}_{\textrm{max}}$ \cellcolor{orange!10}&\cellcolor{orange!10} $\overline{|r|^2}_{\textrm{min}}$\\ \hline
    $3^{rd}$ Moment of mass & $\overline{|r|^3}=\frac{1}{V}\sum_i^V |\vec{r_i}|^3$ \cellcolor{cyan!20}&\cellcolor{orange!10} $\langle\overline{|r|^3}\rangle$ & $\overline{|r|^3}_{\textrm{max}}$ \cellcolor{orange!10}&\cellcolor{orange!10} $\overline{|r|^3}_{\textrm{min}}$\\ \hline
    Stand. var. of mass & $\sigma=\sqrt{\overline{|r|^2}-\overline{|r|}^2}$ \cellcolor{cyan!20}& $\langle\sigma\rangle$ \cellcolor{orange!10}& $\sigma_{\textrm{max}}$ \cellcolor{orange!10}&\cellcolor{orange!10} $\sigma_{\textrm{min}}$\\ \hline
    Skewness of mass & $\mu=(\overline{|r|^3}-3\overline{|r|}\sigma^2-\overline{|r|}^3)/\sigma^3$ \cellcolor{cyan!20}&\cellcolor{orange!10} $\langle\mu\rangle$ & $\mu_{\textrm{max}}$ \cellcolor{orange!10}&\cellcolor{orange!10} $\mu_{\textrm{min}}$\\ \hline    
    Total border length & $B=\sum_j \beta_j$ \cellcolor{cyan!20}& $\langle B \rangle$ \cellcolor{orange!10}& $B_{\textrm{max}}$ \cellcolor{orange!10}&\cellcolor{orange!10} $B_{\textrm{min}}$\\ \hline
    Longest border length & $\beta_l=\max_j(\beta_j)$ \cellcolor{cyan!20}& \cellcolor{gray!25}& \cellcolor{gray!25}& \cellcolor{gray!25}\\ \hline
    Shortest border length & $\beta_s=\min_j(\beta_j)$ \cellcolor{cyan!20}& \cellcolor{gray!25}& \cellcolor{gray!25}& \cellcolor{gray!25}\\ \hline
    $1^{st}$ Moment of border length & $\overline{\beta}=\frac{1}{N}\sum_j \beta_j$ \cellcolor{cyan!20}& $\langle \overline{\beta} \rangle$ \cellcolor{orange!10}& $\overline{\beta}_{\textrm{max}}$ \cellcolor{orange!10}&\cellcolor{orange!10} $\overline{\beta}_{\textrm{min}}$\\ \hline
    $2^{nd}$ Moment of border length & $\overline{\beta^2}=\frac{1}{N}\sum_j \beta_j^2$ \cellcolor{cyan!20}& $\langle \overline{\beta^2} \rangle$ \cellcolor{orange!10}& $\overline{\beta^2}_{\textrm{max}}$ \cellcolor{orange!10}&\cellcolor{orange!10} $\overline{\beta^2}_{\textrm{min}}$\\\hline
    Stand. var. of border length & $\sigma_\beta=\sqrt{\overline{\beta^2}-\overline{\beta}^2}$ \cellcolor{cyan!20}& $\langle \sigma_\beta \rangle$ \cellcolor{orange!10}& $\sigma_{\beta,\textrm{max}}$ \cellcolor{orange!10}&\cellcolor{orange!10} $\sigma_{\beta,\textrm{min}}$\\ \hline
    Semi-major axis & $a$ \cellcolor{cyan!20}& \cellcolor{gray!25} & \cellcolor{gray!25} & \cellcolor{gray!25}\\ \hline
    Semi-minor axis & $b$ \cellcolor{cyan!20}& \cellcolor{gray!25} &  \cellcolor{gray!25} & \cellcolor{gray!25}\\ \hline
    Eccentricity & $e=\sqrt{1-b^2/a^2}$ \cellcolor{cyan!20}& $\langle e \rangle$ \cellcolor{orange!10}& $e_{\textrm{max}}$ \cellcolor{orange!10}&\cellcolor{orange!10} $e_{\textrm{min}}$\\ \hline
    Summed squared residual (between fitted ellipse and boundary pixels) & $R = \sum_i (a x_i^2+ b x_i y_i + c y_i^2 + d x_i + ey_i +f)^2$ \cellcolor{cyan!20}& $\langle R \rangle$ \cellcolor{orange!10}& $R_{\textrm{max}}$ \cellcolor{orange!10}&\cellcolor{orange!10} $R_{\textrm{min}}$\\ \hline
       Parallel alignment & $\Gamma_{\parallel}=\frac{1}{N}\sum_j (1-\frac{bb_j}{aa_j})|\cos{(\theta-\theta_j)|}$ \cellcolor{cyan!20}&\cellcolor{orange!10} $\langle \Gamma_{\parallel} \rangle$ & $\Gamma_{\parallel,\textrm{max}}$ \cellcolor{orange!10}&\cellcolor{orange!10} $\Gamma_{\parallel,\textrm{min}}$ \\ \hline
    Perpendicular alignment & $\Gamma_{\perp}=\frac{1}{N}\sum_j (1-\frac{bb_j}{aa_j})|\sin{(\theta-\theta_j)|}$ \cellcolor{cyan!20}&\cellcolor{orange!10} $\langle \Gamma_{\perp} \rangle$ & $\Gamma_{\perp,\textrm{max}}$\cellcolor{orange!10} &\cellcolor{orange!10} $\Gamma_{\perp,\textrm{min}}$ \\ \hline
    Parallel front-end alignment & $\Gamma_{\parallel,\textrm{FE}} =   \frac{1}{N}\sum_j (1-\frac{bb_j}{aa_j})|\cos{(\theta-\theta_j)}\cos{(\theta-\phi_j)}|$ \cellcolor{cyan!20}& $\langle \Gamma_{\parallel,\textrm{FE}} \rangle$ \cellcolor{orange!10}& $\Gamma_{\parallel,\textrm{FE},\textrm{max}}$ \cellcolor{orange!10}&\cellcolor{orange!10} $\Gamma_{\parallel,\textrm{FE},\textrm{min}}$ \\ \hline
    Perpendicular front-end alignment & $\Gamma_{\perp,\textrm{FE}} =   \frac{1}{N}\sum_j (1-\frac{b_j}{a_j})|\sin{(\theta-\theta_j)}\cos{(\theta-\phi_j)}|$\cellcolor{cyan!20}&\cellcolor{orange!10} $\langle \Gamma_{\perp,\textrm{FE}} \rangle$ & $\Gamma_{\perp,\textrm{FE},\textrm{max}}$ \cellcolor{orange!10}&\cellcolor{orange!10} $\Gamma_{\perp,\textrm{FE},\textrm{min}}$ \\ \hline
    Parallel side alignment & $\Gamma_{\parallel,\textrm{side}} =   \frac{1}{N}\sum_j (1-\frac{bb_j}{aa_j})|\cos{(\theta-\theta_j)}\sin{(\theta-\phi_j)}|$ \cellcolor{cyan!20}&\cellcolor{orange!10} $ \langle \Gamma_{\parallel,\textrm{side}} \rangle $ & $\Gamma_{\parallel,\textrm{side},\textrm{max}}$ \cellcolor{orange!10}&\cellcolor{orange!10} $\Gamma_{\parallel,\textrm{side},\textrm{min}}$\\ \hline
    Perpendicular side alignment & $\Gamma_{\perp,\textrm{side}} =   \frac{1}{N}\sum_j (1-\frac{bb_j}{aa_j})|\sin{(\theta-\theta_j)}\sin{(\theta-\phi_j)}|$ \cellcolor{cyan!20}&\cellcolor{orange!10} $ \langle \Gamma_{\perp,\textrm{side}} \rangle $ \cellcolor{orange!10}&\cellcolor{orange!10} $\Gamma_{\perp,\textrm{side},\textrm{max}}$ &\cellcolor{orange!10}$\Gamma_{\perp,\textrm{side},\textrm{min}}$\\ \hline
    $1^{st}$ Moment of neighbour distance & $\overline{|R|}=\frac{1}{N}\sum_j |\vec{R_j}|$ \cellcolor{green!10}& $\langle \overline{|R|} \rangle$\cellcolor{violet!30}& $\overline{|R|}_\textrm{max}$ \cellcolor{violet!30}&\cellcolor{violet!30} $\overline{|R|}_\textrm{min}$\\ \hline
    $2^{nd}$ Moment of neighbour distance & $\overline{|R|^2}=\frac{1}{N}\sum_j|\vec{R_j}|^2$ \cellcolor{green!10}&\cellcolor{violet!30}$\langle \overline{|R|^2} \rangle$& $\overline{|R|^2}_\textrm{max}$ \cellcolor{violet!30}&\cellcolor{violet!30} $\overline{|R|^2}_\textrm{min}$\\ \hline
    Standard variation of neighbour distance & $\sigma_R=\sqrt{\overline{|R|^2}-\overline{|R|}^2}$ \cellcolor{green!10}&\cellcolor{violet!30} $ \langle \sigma_R \rangle $ & $ \sigma_{R,\textrm{max}}$ \cellcolor{violet!30}&\cellcolor{violet!30} $\sigma_{R,\textrm{min}}$\\ \hline
    Bond order parameters, for $n=2,...,12$ & $\Psi_n=|\frac{1}{N}\sum_j \cos{(n\phi_j)}^2 + \sin{(n\phi_j)}^2|$ \cellcolor{green!10}&\cellcolor{violet!30} $\langle \Psi_n \rangle$ & $ \Psi_{n,\textrm{max}}$ \cellcolor{violet!30}&\cellcolor{violet!30} $\Psi_{n,\textrm{min}}$ \\ \hline
    Bond order parameter for $n=N$ & $\Psi_N=|\frac{1}{N}\sum_j \cos{(N\phi_j)}^2 + \sin{(N\phi_j)}^2|$ \cellcolor{green!10}&\cellcolor{violet!30} $\langle \Psi_N \rangle$ & $ \Psi_{N,\textrm{max}}$ \cellcolor{violet!30}&\cellcolor{violet!30} $\Psi_{N,\textrm{min}}$\\ \hline
    \end{tabular}
\end{table*}

Rather than using simulation snapshots as input features, we extract single-cell features from each snapshot to use as input for our simple machine-learning model. This approach is preferred because it provides interpretable results. We employ a total of 145 possible features as input for our machine-learning model, categorising them into structure and shape features. The comprehensive list of these features and the formulas used for their computation are shown in Tab.~\ref{tab:features}.
Structural features are derived from the centres of mass (COM) of cells and are based solely on properties akin to local structural metrics commonly used for dense, disordered particle systems. These features encompass bond order parameters $\psi_n$ with $n=2,\dots,12$~\cite{Mickel2013}, along with the first and second moment of the neighbour distance and its standard deviation.
The single-cell shape features, instead, are computed based on the pixels that constitute each cell. These geometric features include cell size, border length, semi-minor and semi-major axes, parallel and perpendicular alignment, number of neighbouring cells (calculated for each cell to determine how many other cells are adjacent to it), and eccentricity. The eccentricity is determined by fitting cells with an ellipse using a least squares approach~\cite{pilu1996ellipse}.

For both shape and structural features, we further divide these types into two categories: local features, derived from information about individual cells, and non-local features, which depend on the properties of a cell's neighbours. Since active cells tend to deform their neighbourhoods more significantly, examining non-local features provides additional insights. These non-local properties include neighbour averages, and maximum and minimum distances between the centres of mass of a cell and its neighbouring cells. Similar to Ref.~\cite{Wang_2020_AnisotropyLinkShapeToFlow}, local shape alignment between neighbouring cells has also been included. Table~\ref{tab:features} illustrates local shape features in blue, local structure features in green, non-local shape features in orange, and non-local structure features in violet. 

Note that the list of features used here is by no means complete. Depending on the specific biological situations, other features could be relevant as well. For example, individual human bone marrow stromal cells (hBMSCs) exhibit strong surface curvature, which can also be a relevant shape characteristic to include~\cite{Chen_MLtoidentifycellshapephenotypes}. These features have not been included here, since the cells in the simulations do not exhibit strong curvature. Moreover, it is worth noting that additional radial and angular descriptors can be incorporated into the structural features, as outlined in Ref.~\cite{Boattini2021}. However, we choose to focus on a simpler approach for computational efficiency~\cite{Janzen_2023_DeathAlive} and because, as will be explained in the results section, our approach, though simple, is robust and provides sufficiently accurate results.

\subsection{Feature selection}
To achieve optimal performance and gain physical insight from the ML predictions, we evaluate the importance of input features using three different approaches: manually removing some features, Shapley Additive Explanation (SHAP), and Principal Component Analysis (PCA).
The first approach involves training seven different neural networks: one with all the features and the remaining six with subsets of the entire dataset. These subsets include shape features (both local and non-local), local shape features, non-local shape features, structural features (both local and non-local), local structural features, and non-local structural features. After training, we evaluate which neural network achieves the highest accuracy on the test set.

Our second approach involves using SHAP~\cite{lundberg2017unified} to determine the relative contribution of each feature to the prediction. In essence, the SHAP explanation method computes Shapley values by integrating concepts from cooperative game theory. The objective of this analysis is to distribute the total payoff among players, considering the significance of their contributions to the final outcome. In this context, the feature values act as players, the model represents the coalition, and the payoff corresponds to the model's prediction.

Lastly, our third approach involves applying PCA~\cite{jolliffe2002principal} on our dataset, including shape and structural features. PCA is a valuable tool for condensing multidimensional data with correlated variables into new variables, representing linear combinations of the original ones. Essentially, PCA serves as a method to reduce the dimensionality of high-dimensional data.  By identifying the features with significant variances, we can reveal the inherent characteristics within our dataset. The first component corresponds to the projection axis that maximises variance in a particular direction, whereas the second principal component represents an orthogonal projection axis that maximises variance along the subsequent leading direction. This iterative process can be continued to identify additional components.

\section{Results and discussion} \label{sec:res}
\subsection{Distinguishing motile and non-motile cells} \label{sec:res-passive}
\begin{figure}
    \centering
    \includegraphics[width=\linewidth]{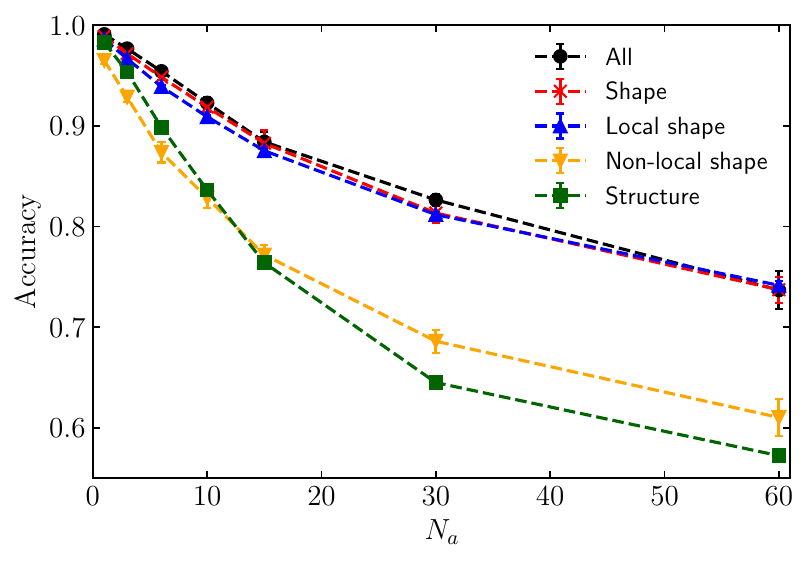}
    \caption{Accuracy as a function of the number of active particles $N_a$, with $\kappa_p=0$ and $\kappa_a=1500$. The black dots, red stars, blue triangles, inverted triangles and green squares correspond to a neural network trained on all the 145 features, all the shape features, local shape features, non-local shape features, and structural features respectively. The lines are used as a guide to the eye. }
    \label{fig:accuracy_local}
\end{figure}

Let us first focus on the situation in which passive cells are non-motile ($\kappa_p~=~0$) and the active cells are highly motile ($\kappa_a~=~1500$).
Figure \ref{fig:accuracy_local} shows the accuracy as a function of the number of active cells, $N_a$. The neural network is trained with different feature configurations, encompassing either all 145 features (black dots), all shape features only (both local and non-local, represented by red stars), solely local shape features (blue triangles), exclusively non-local shape features (orange inverted triangles), and only structural features (both local and non-local, represented by green squares). All five curves produce comparable accuracies, approaching unity when a single active cell moves through a non-motile confluent layer. This result is expected given that the active cell, characterised by a more elongated shape compared to the passive cells, is the only one present, making it easily distinguishable even to the naked eye (see Fig.~\ref{fig:overview_all}). Across all four datasets, as the number of active particles increases, the accuracy decreases. Nevertheless, when the neural network is trained with all features, all shape features, or only local shape features, the accuracy remains above 0.7, suggesting that the algorithm can classify the cell's motility with reasonable accuracy. On the contrary, employing structural features alone results in significantly lower accuracy compared to the full dataset or the shape features.  Hence, the shape of individual cells contains a substantial amount of information regarding the cell's motility. 

To gain a deeper understanding of the importance of shape features, we have further subdivided the shape features into local features (single-cell information) and non-local features (information from neighbouring cells). As shown in Fig.~\ref{fig:accuracy_local}, it is noteworthy that relying solely on local shape features predicts the correct cell phenotype with almost the same accuracy as using the full set of features. 

Lastly, we perform analyses using both SHAP and PCA. Both reveal that the list of important features is not limited to local shape features but rather encompasses a combination of the four feature groups: shape (both local and non-local) and structural (both local and non-local) features. Retraining the neural network with the features selected by SHAP or with the principal components obtained from the PCA yields an accuracy almost identical to that obtained with a neural network trained with all features (see Supplementary material). As shown in the Supplementary material, these analyses indicate that features related to neighbour distance (e.g., standard deviation of neighbour distance) are often the most important ones. Since the neighbour distance-related features are indirectly connected to the shape of the cells, it is perhaps not surprising that these analyses identify these features as the most important ones. 

Although SHAP and PCA reveal that the most important features are a combination of both shape and structure, the list of relevant features selected by these machine-learning approaches changes with $N_a$, making these analyses less computationally efficient. This inefficiency arises from the need to repeat these analyses (SHAP or PCA) for each specific configuration to obtain this list of most important features. Therefore, we can conclude that our simpler approach of selecting only shape features is sufficient for achieving reasonable accuracy and is the most robust, consistently yielding results almost identical to those obtained using all features, regardless of $N_a$.

\subsection{Distinguishing cells with different motility} \label{sec:res-bothactive}
In the previous section, we have shown that a neural network trained with local shape features can correctly predict the cell’s motility when the passive cells are non-motile ($\kappa_p~=~0$) and the active cells are significantly more motile ($\kappa_a~=~1500$). However, in realistic heterogeneous biological tissues, cells are expected to have different but finite degrees of  motility~\cite{Bi_Motility_JammingTransition,cerchiari2015strategy,ravindran2019cross}. To study a system that more closely resembles actual biological systems, albeit still simplified, we focus on a binary mixture of active and passive cells where the passive cells are also motile, with a fixed motility $\kappa_p~=~150$. However, their motility remains lower than that of active cells ($\kappa_p~<~\kappa_a$), where $\kappa_a$ is varied between $300$ and $1500$.

\begin{figure}
    \centering
    \includegraphics[width=\linewidth]{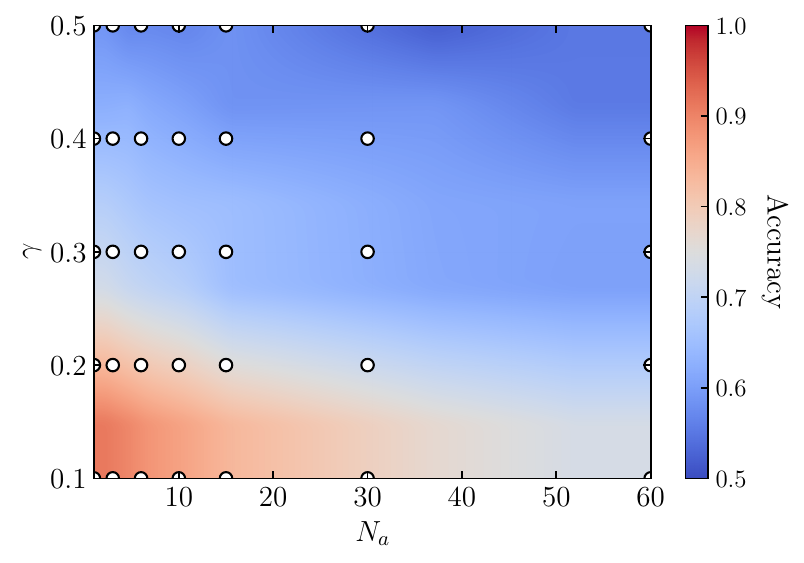}
    \caption{Accuracy map of a neural network trained on local shape features in the $(\gamma, N_a)$-plane, where $N_a$ ranges from 1 to 60 and $\gamma=\kappa_p/\kappa_a$, with $\kappa_p=150$ and $300 \le \kappa_a \le 1500$.}
    \label{fig:accuracy_heatmap_all_local}
\end{figure}

Following a similar approach as in the previous section, we train a neural network for each dataset using static properties, as introduced in Sec.~\ref{sec:input_features}. Here, each dataset corresponds to a distinct ratio between passive and active cell motility, denoted as $\gamma=\kappa_p/\kappa_a$, along with the number of active cells $N_a$. 
Figure~\ref{fig:accuracy_heatmap_all_local} shows the accuracy within the $(\gamma,~N_a)$-plane for a neural network trained with only local shape features. Consistent with the results observed for non-motile passive cells, the neural network exclusively trained on local shape features has nearly identical accuracy compared to the one trained with all 145 features (see Supplementary material). This figure shows that when the number of active cells $N_a$ is low, and the ratio between cell motility $\gamma$ is small, indicating a substantial difference between active and passive cells, the model can accurately classify the cell motility. 

While the machine learning model relies on individual static images, our numerical CPM simulations also enable the explicit tracking of the emergent dynamics. Notably, we find that our machine learning model tends to fail only when the emergent dynamics, specifically the long-time diffusion coefficients, of active and passive cells are very similar. These findings align with those presented in Ref.~\cite{Janzen_2023_DeathAlive}, where it was shown that in an active-passive mixture of spherical, rigid particles, a machine learning model can correctly classify particle types when the number of active particles is low, and the activity is high.

Finally, invoking a SHAP analysis or PCA, we achieve accurate predictions using only the most important SHAP- or PCA-selected features (see Supplementary material). Similarly to the previous section, where passive cells are non-motile, we observe that the most important features identified by these analyses are a combination of shape (both local and non-local) and structural (both local and non-local) features. While this feature list remains consistent for fixed $\gamma>0$ and different $N_a$, it varies for different $\gamma$. Consequently, as discussed in the previous section, this approach is less computationally efficient compared to the case of using local shape features, which yields accurate results for different configurations. 

In summary, our findings indicate that a machine-learning model can accurately classify cell motility when the number of active cells is low, and the motility of active cells significantly surpasses that of passive cells.
Comparable to the case in which the passive cells are non-motile, accurate predictions can be achieved using local shape features alone.

\subsection{Generalisation of the model} \label{sec:res-generalize}
We now aim to assess how effectively our machine-learning model generalises to different data featuring either a distinct number of motile cells or varying motility.
First, we explore the generalisation capability of the machine learning model when the passive cells are non-motile ($\kappa_p=0$), and the number of active cells $N_a$ varies. In the previous section, we established that a model trained exclusively on local shape features achieves an accuracy almost identical to that of a model trained with all 145 features. Therefore, in the remainder of this paper, we present the results from neural networks trained exclusively on local shape features.

\begin{figure}
    \centering
    \includegraphics[width=\linewidth]{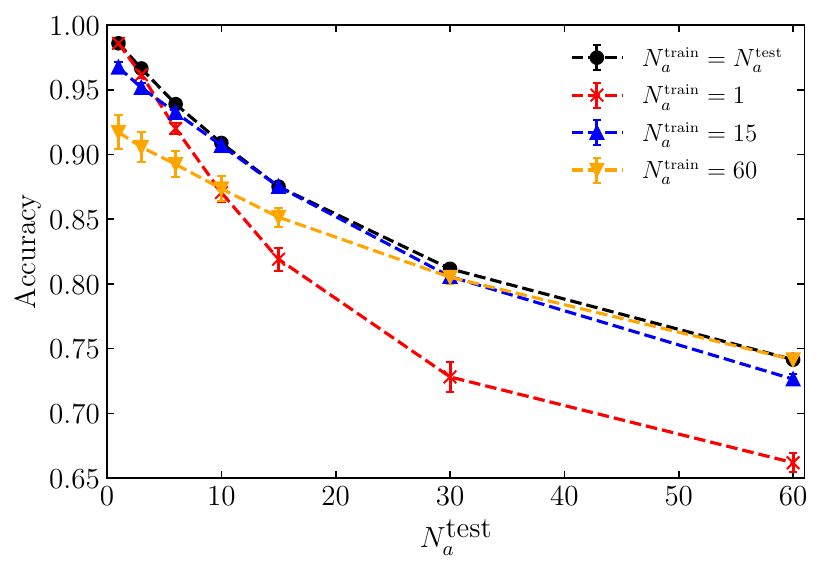}
    \caption{Accuracy as a function of the number of active cells $N_a=N_a^{test}$, with $\kappa_p=0$ and $\kappa_a=1500$. The black dots represent accuracy obtained from individual models, each trained using $N_a^{train}=N_a^{test}$. The red stars, blue triangles, and orange inverted triangles represent accuracy obtained from a single model trained with data for $N_a=1$, $N_a=15$ or $N_a=60$, respectively. Each neural network is trained exclusively with local shape features.}
    \label{fig:accuracy_phi}
\end{figure}

In Fig.~\ref{fig:accuracy_phi}, we compare the accuracy obtained when the ML model is trained and tested on a singular specific value of the number of active cells $N_a$ (black dots), with the accuracy of models trained with $N_a=1$ (red stars), $N_a=15$ (blue triangles), or $N_a=60$ (orange inverted triangles).
As expected, this figure shows that the model performs best when trained and tested on a singular, specific value of $N_a$. Nevertheless, all four curves yield an accuracy surpassing $0.7$, suggesting that a single model trained at a fixed $N_a$ can provide accurate predictions for unseen parameter regions. 

Figure~\ref{fig:accuracy_phi} reveals that the ML model trained with an intermediate number of active cells, $N_a=15$, yields nearly identical results compared to the model trained and tested on a single, specific value of $N_a$. This model is the most effective across the entire range of $N_a$. We attribute this to the fact that a system with an intermediate number of active cells shares similarities with both low and high numbers of active cells, contributing to its robust performance.
Additionally, while the model trained on one active cell generalises better to different data associated with a small number of active cells ($N_a~<~10$), the model trained on $N_a=60$ generalises better to different data corresponding to a high number of active cells ($N_a > 30$). This discrepancy arises from the distinctive system structures between scenarios with only one active cell and those with a substantial number, respectively.

Lastly, we explore whether the machine learning approach can generalise to a different data set when the number of active cells is constant, and the ratio between cell motilities $\gamma$ varies. For each value of $N_a$, we train four distinct models: one with $\gamma=0$ (where $\kappa_p=0$ and $\kappa_a=1500$), another with $\gamma=0.1$ (where $\kappa_p=150$ and $\kappa_a=1500$), a third with $\gamma=0.2$ (where $\kappa_p=150$ and $\kappa_a=750$), and the last one with $\gamma=0.4$ (where $\kappa_p=150$ and $\kappa_a=375$). Subsequently, each of these four models is tested with the local shape features corresponding to the dataset with a fixed ratio $\gamma=0.1$.

\begin{figure}
    \centering
    \includegraphics[width=\linewidth]{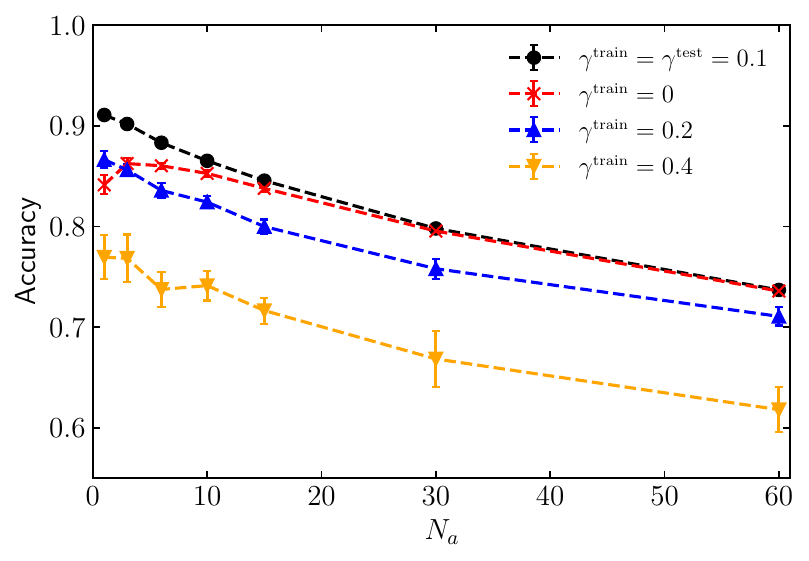}
    \caption{Accuracy as a function of the number of active cells $N_a$, for neural networks solely trained on local shape features. The black dots represent accuracy obtained from individual models, each trained using $\gamma^{train}=\gamma^{test}=0.1$. The red stars, blue triangles, and orange inverted triangles represent accuracy obtained from models trained with data for $\gamma=0$, $\gamma=0.2$ or $\gamma=0.4$, respectively. The accuracy for each of these lines corresponds to the neural network tested on $\gamma=0.1$.}
    \label{fig:accuracy_N}
\end{figure}

Figure~\ref{fig:accuracy_N} demonstrates that, as expected, the highest performance is achieved by a model trained and tested on the identical ratio between cell motility, $\gamma=0.1$ (black dots). Additionally, the figure shows that a model trained on $\gamma=0$ (represented by red stars) yields an accuracy nearly indistinguishable from the model trained and tested on $\gamma=0.1$ when the number of active particles is high ($N_a>10$). 
In both datasets corresponding to $\gamma=0$ and $0.1$, the motility of the active cells remains constant. Consequently, the model exhibits effective generalisation within this parameter range, even if is trained on data associated with different passive motility. This robust generalisation can be attributed to the similarity in behaviour between the two systems, given the abundant active particles sharing the same motility.

When the model is trained on $\gamma=0.2$ and tested on $\gamma=0.1$ (blue triangles), the accuracy is always lower than that of the model trained and tested on $\gamma=0.1$. Nonetheless, the accuracy is consistently higher than $0.7$, indicating that this model can reasonably generalise unseen data. 
Lastly, when the model is trained on $\gamma=0.4$ (orange inverted triangles), the accuracy significantly diminishes compared to the model trained and tested on $\gamma=0.1$. Furthermore, the accuracy drops below $0.7$ as the number of active cells increases ($N_a>20$). These results indicate that a decrease in the motility of active cells corresponds to a lower predictive power of the model when tested on unseen data.

In summary, we find that the generalisation capability of our machine-learning approach to different unseen data is reasonable. The model is capable of making fairly accurate predictions when the number of active cells is unknown, but its predictive power diminishes when the motility of the active cells in the training and testing sets are significantly different.

\section{Conclusions} \label{sec:conclusion}
This study establishes proof-of-concept for discriminating between highly motile (active) cells and less motile or non-motile (passive) cells within a heterogeneous confluent cell layer, using only static information of a cell's instantaneous shape and structural environment.
Our results show that a simple machine-learning model trained on local, single-cell shape features alone can predict the cellular motility phenotype with reasonably good accuracy, and excels especially when the fraction of active cells is low and their motility is significantly higher than that of passive cells~\cite{Janzen_2023_DeathAlive}.  
While prior studies have also highlighted the importance of cell shape and morphology~\cite{Kim_2023_PredictionStemCellState_DeepLearning,Lyons_CellShape_correlates_metastaticrisk,DOrazio_DecipheringCancerCellBehavior, Bi_rigidity_nonmotile}, most notably in strongly anisotropic tissues, other measures, such as the alignment between cells, may be necessary for a fully accurate prediction~\cite{Wang_2020_AnisotropyLinkShapeToFlow}. 

Common limitations of machine-learning approaches are that they may generalise poorly to unseen data, and that they may offer limited physical insight. We find that our model exhibits reasonably good generalisation when the number of active cells or the motility ratio is unknown, provided that the motility strengths in the training and testing sets do not differ greatly. This reaffirms that the power of machine-learning methods relies heavily on the use of a sufficiently diverse data set. Additionally, to gain some physical insight from our machine-learning predictions, we have employed three different methods to assess the importance of the various input features. Of these, the analyses based on SHAP and PCA reveal that there is not a universal list of most important static features: in general, the most important features combine cellular shape and structural characteristics, and the list varies with different heterogeneity settings ($N_a$). Nonetheless, if we restrict the data set to local single-cell shape features alone, we find that this simple approach leads to remarkably robust predictions across the different settings studied in this work. This suggests that the full list of structural input features may contain some redundancies. Importantly, it also allows us to conclude that a cell's instantaneous shape, though not perfect, can serve as a remarkably useful informant on a cell's phenotype.

Our work, which establishes a morphodynamic link for individual cells, is complementary to recent research on morphodynamic links at the collective cell level. In particular, previous studies have demonstrated that the average cell shape within confluent tissue can be used as a static order parameter for emergent, collective cell jamming and unjamming dynamics \cite{Bi_rigidity_nonmotile,Bi_Motility_JammingTransition, Chiang_GlassyInCPM, Czajkowski_HydrodynamicsShapeDrivenTransition, Devanny_Jamming_CPM, Atia2018_geometricconstraint, Park2015, Kim_2020_UnjammingMigration_CancerCell}. 
By integrating these insights, our work not only reinforces the significance of cell shape in understanding collective behaviour, but it also provides a more nuanced perspective on how intrinsic single-cell properties are coupled to a cell's morphology. 
In view of the simplicity, performance, and computational efficiency of our machine-learning approach, we envision that a similar approach may also be of use in the analysis of experimental cell data, particularly for diagnostic tasks such as estimating the progression of a partial or full EMT in tumours or tissues. 

\section*{Acknowledgements}
QJSB and LMCJ thank the Dutch Research Council (NWO) for financial support through the ENW-XL project ``Active Matter Physics of Collective Metastasis" (OCENW.GROOT.2019.022).

\bibliographystyle{apsrev4-1} 
\bibliography{./references}

\end{document}